\documentstyle[prl,aps,multicol,epsf,rotate]{revtex}
\begin{document}
\draft
\preprint{XXXX}
\title{Upper critical dimension, dynamic
exponent and scaling functions in the
mode-coupling theory for the
Kardar-Parisi-Zhang equation}
\author{Francesca Colaiori and M. A. Moore}
\address{Department of Physics and Astronomy, University of Manchester,
Manchester, M13 9PL, United Kingdom}
\date{\today}
\maketitle
\begin{abstract}
We study the mode-coupling approximation for
the KPZ equation in the strong coupling regime.
By constructing an ansatz consistent with the asymptotic forms of
the correlation and response functions we determine the upper critical
dimension $d_c=4$, and the expansion $z=2-(d-4)/4+O((4-d)^2)$ around $d_c$.
We find the exact $z=3/2$ value in $d=1$, and estimate the values
$z\simeq1.62$, $z\simeq 1.78$, in $d=2,3$.
The result $d_c=4$ and the expansion around $d_c$
are very robust and can be derived just from a mild
assumption on the relative scale on which the
response and correlation functions vary as $z$ approaches $2$.
\end{abstract}
\begin{multicols}{2}
The Kardar-Parisi-Zhang (KPZ) \cite{KPZ} equation is a simple nonlinear
Langevin equation, proposed in 1986 as a coarse-grained
description of a variety of growth processes.
Due to its connection to many other important physical problems
(such as the Burgers equation \cite{For}, directed polymers in a random
medium \cite{Kar,Der,Par}, etc.), it has attracted
much attention.
When applied to a growing interface, described by a
single valued height function $h({\bf x},t)$ on a $d$-dimensional
substrate ${\bf x}$ the KPZ equation is:
\begin{equation}
\partial_t h({\bf x},t)=
\nu \nabla^2 h +\frac{\lambda}{2}(\nabla h)^2+\eta({\bf x},t),
\label{KPZ}
\end{equation}
where the first term represents the surface tension forces which tend to
smooth the interface, the second describes the non-linear growth locally
normal to the surface, and the last is a noise term intended to
mimic the stochastic nature of the growth process \cite{rev}.
We choose the noise to be Gaussian, with zero mean and second moment
$\langle \eta({\bf x},t)\eta({\bf x'},t')\rangle=
2D\delta^{d}({\bf x}-{\bf x'})\delta(t-t').$
The steady state interface profile is usually described in terms
of the roughness:
$w=\langle h^2({\bf x},t)\rangle- \langle h({\bf x},t)\rangle^2 $
which for a system of size $L$ behaves like $L^{\chi} f(t/L^z)$,
where $f(x)\rightarrow const$ as $x\rightarrow \infty$
and $f(x) \sim x^{\chi/z}$ as $x \rightarrow 0$, so that $w$
grows with time like $t^{\chi/z}$ until it saturate to $L^{\chi}$
when $t\sim L^z$. $\chi$ and $z$ are the roughness and dynamic
exponent respectively.

For $d>2$ there are two distinct regimes, separated by a critical value
of the nonlinearity coefficient. In the weak coupling regime
($\lambda < \lambda_c$) the non-linear term is irrelevant
and the behavior is governed by the $\lambda=0$ fixed point. The
linear Edward-Wilkinson equation is recovered, for which the
exponents are known exactly $\chi=(2-d)/2$ and $z=2$.
The more challenging strong coupling regime ($\lambda > \lambda_c$),
where the non-linear term is relevant is characterised by anomalous
exponents, which are the subject of this paper. From the Galilean invariance
\cite{For}
(the invariance of equation (\ref{KPZ}) under an infinitesimal
tilting of the surface) one can derive the relation
$\chi+z=2$, which leaves just one independent exponent. For the special case
$d=1$, the existence of
a fluctuation-dissipation theorem leads to the exact result
$\chi=1/2$, $z=3/2$.

While we have a satisfactory understanding of the KPZ equation
in $d=1$ \cite{K,Hwa} and on the Bethe lattice \cite{Der}, its behavior
in the strong coupling regime
when $d>1$ is controversial. Such results that are known derive mainly
from numerical simulations \cite{Num} of discrete models
which belong to the KPZ universality class \cite{Mar}, and
on several promising but far from conclusive analytical approaches
\cite{Sch,Bou,Do,Mo,Blu,Bha,La,Wie}.
Most efforts are oriented towards the determination
of $z$ as a function of $d$ for $d>1$, and to understanding
whether there exists an upper critical dimension $d_c$ above
which $z=2$: some analytical arguments suggest
the existence of a finite $d_c$ \cite{Mo,Blu,Bha,La,Wie,Hh}, while
numerical studies
\cite{Num} and real space methods \cite{Cla} find no evidence of a finite
upper critical dimension. It is our contention that $d_c =4$, and that the
numerical simulations all fail to see evidence for an upper critical
dimension because in high dimensions, ($d\geq4$), they have not been run for
sufficiently long time for the roughness $w$ to be in the scaling regime.

We take a self-consistent approach, the mode-coupling (MC) approximation
\cite{For}, where in the diagrammatic expansion for the correlation
and response function only diagrams which do not renormalize the three point
vertex $\lambda$ are retained.  The MC approximation has
been remarkably successful; it reproduces the exact values
of the exponents in $d=1$ and  gives an accurate value for $z$ in
$d=2$. Furthermore the MC equations have been shown
to arise from the large $N$-limit of an generalised $N$-component KPZ
\cite{Do}, which allows in principle a systematic approach to the
theory beyond MC, by expanding in $1/N$.

Because of its relative simplicity and its seeming accuracy, there have been
many studies of the mode-coupling approach to KPZ:
Hwa, Frey and Tauber \cite{Hwa} studied the $d=1$ case, where it gives the
exact result $z=3/2$, and solved numerically for the scaling
functions. Bouchaud and Cates \cite{Bou} gave an approximate
analytical solution assuming simple exponential relaxation for
each mode. Doherty et al. \cite{Do} used instead an ansatz based on the form
of the scaling functions in $d=1$. Both \cite{Bou,Do} found a finite
upper critical dimension, with the same  value $d_c\simeq 3.6$.
Moore et al. \cite{Mo} found an explicit solution of the MC equations
for $d>d_{c}=4$ with $z=2$.
(Calculations outside the MC framework  which also give $d_c=4$ include a
functional RG calculation \cite{Hh}, renormalization group
arguments \cite{La,Wie}, and an expansion
around $z=2$ \cite{Bha}. None of these give such
compelling arguments for $d_c=4$ to have won general acceptance either 
for that value for the upper critical dimension or even for the existence of 
an upper critical dimension).
Tu \cite{Tu} attempted the formidable task of numerically solving the mode
coupling equations by direct
integration, and found no evidence for an upper critical dimension. We shall
comment later on why with his particular numerical technique,  problems in
locating the upper critical dimension would arise. We believe that one of
the reasons why there is a reluctance to accept that there is an upper
critical dimension is that earlier approximate solutions of the MC equations
yielded unconvincing non-integer values for the upper critical dimension
(such as $3.6...$) and that the approach  which did give an integer value of
four for $d_c$, \cite{Mo}, did not, however, provide any insights as to the
form of the solution in the physically accessible regime, $d<4$. In this
paper we shall give an (approximate) analytical treatment of the MC
equations which predicts that $d_c$ is four, gives z correct to $O(4-d)$ and
provides some insights into the form of the scaling functions.

The correlation and response function are defined in Fourier space by
\begin{eqnarray}
&C({\bf k},\omega)=\langle h({\bf k},\omega) h^{*}({\bf k},\omega) \rangle,
\nonumber
\\
&G({\bf k},\omega)=\delta^{-d}({\bf k}-{\bf k}^{'})\delta^{-1}
(\omega-\omega^{'})
\langle
\frac{\partial h({\bf k},\omega)}{\partial \eta({\bf k}^{'},\omega^{'})}
\rangle ,
\nonumber
\end{eqnarray}
\noindent
where $\langle \cdot \rangle$ indicate an average over $\eta$.
In the mode coupling approximation, the correlation and response
functions are the solutions of two coupled equations,
\end{multicols}
\begin{eqnarray}
&
G^{-1}({\bf k},\omega)=G^{-1}_0({\bf k},\omega)+\lambda^2
\int \frac{d\Omega}{2 \pi} \int \frac{d^dq}{(2 \pi)^d}
\left[{\bf q} \cdot ({\bf k}-{\bf q})\right]\left[{\bf q} \cdot {\bf
k}\right]
G({\bf k}-{\bf q},\omega - \Omega) C({\bf q},\Omega) \,,
\label{mc1}
\\&
C({\bf k},\omega)=C_0({\bf k},\omega)+\frac{\lambda^2}{2}
\mid G({\bf k},\omega)\mid^2
\int \frac{d\Omega}{2 \pi} \int \frac{d^dq}{(2 \pi)^d}
\left[{\bf q} \cdot ({\bf k}-{\bf q})\right]^2
C({\bf k}-{\bf q},\omega - \Omega) C({\bf q},\Omega) \,,
\label{mc2}
\end{eqnarray}
%\begin{multicols}{2}
\noindent
where $G_0({\bf k},\omega)=(\nu k^2 - i \omega)^{-1}$ is the bare
response function, and $C_0({\bf k},\omega)=2 D \mid G({\bf
k},\omega)\mid^2$.
In the strong coupling limit,
$G({\bf k},\omega)$ and $C({\bf k},\omega)$ take the following
scaling forms
\begin{eqnarray}
&
G({\bf k},\omega)=k^{-z}g\left( \omega/k^{z}\right) \,,
\nonumber
\\&
C({\bf k},\omega)=k^{-(2 \chi+d+z)}n\left( \omega/k^{z}\right) \,,
\nonumber
\end{eqnarray}
\noindent
and Eqs. (\ref{mc1}) and  (\ref{mc2})
translate into the following coupled
equations for the scaling functions $n(x)$ and $g(x)$:
\begin{eqnarray}
&
g^{-1}(x)=-i x +I_{1}(x) \,,
\label{g}
\\
&
n(x)=\mid g(x)\mid^{2}I_{2}(x) \,,
\label{n}
\end{eqnarray}
where $x=\omega/k^z$ and $I_1(x)$ and $I_2(x)$ are given by
%\end{multicols}
\begin{eqnarray}
&
I_1(x)=P\int_{0}^{\pi}d\theta \sin^{d-2}\theta
\int_{0}^{\infty}dq\cos\theta (\cos\theta-q)q^{2z-3}r^{-z}
\int_{-\infty}^{\infty}dy \,g\left(\frac{x-q^{z}y}{r^{z}}\right)n(y),
\nonumber
\\
&
I_2(x)=\frac{P}{2}\int_{0}^{\pi}d\theta \sin^{d-2}\theta
\int_{0}^{\infty}dq(\cos\theta-q)^2 q^{2z-3}r^{-(d+4-z)}
\int_{-\infty}^{\infty}dy \,n\left(\frac{x-q^{z}y}{r^{z}}\right)n(y),
\nonumber
\end{eqnarray}
\begin{multicols}{2}
\noindent
with
$P\!=\!\lambda^2/(2^d\Gamma(\frac{d-1}{2})\pi^{(d+3)/2})$,
$r^2=1+q^2-2q\cos\theta$.

All the approximate analytical approaches start by making an ansatz on the
form of the scaling
functions $n(x)$ and $g(x)$. The relation $z=z(d)$ is then obtained
requiring consistency of Eqs. (\ref{g}) and (\ref{n}) 
on matching both sides at an arbitrarily chosen value of $x$.
Due to the non-locality of Eqs. (\ref{g}) and (\ref{n}), the matching
condition depends on the form of the functions $n$ and $g$
for all $x$, so the ansatz need to be reliable for all $x$.
Previous ansatzes  were unsatisfactory, since (as already pointed out by
Tu \cite{Tu}), they did not satisfy the large $x$ asymptotic forms
\begin{equation}
n(x)\sim x^{-1-\beta/z}, \,\, g_R(x)\sim x^{-1-2/z},\,\, g_I(x)\rightarrow 
x^{-1},
\label{asymp}
\end{equation}
where $\beta=d+4-2z$ and $g(x)=g_R(x)+ig_I(x)$. \\
We will match the left and right hand sides of both Eqs. (\ref{g}) and Eq.
(\ref{n}) as $x\rightarrow \infty$
using an ansatz that capture this large $x$ asymptotic behavior. It is by
doing this that we are able to obtain $z$ correct to $O(4-d)$.
It is convenient to first write Eqs. (\ref{g}) and (\ref{n})
in Fourier space:
\begin{equation}
{\widehat{\,\,\,\frac{g_R}{\mid g\mid^2}\,\,\,}}(p)=\widehat{I_1}(p) \,,
\label{ghat}
\end{equation}
\begin{equation}
\widehat{\,\,\,\frac{n}{\mid g\mid^2}\,\,\,}(p)=\widehat{I_2}(p) \,,
\label{nhat}
\end{equation}
where $\widehat{I_1}$ is the Fourier transform of the real part of $I_1$
and $\widehat{I_2}$ is the Fourier transform of $I_2$
\end{multicols}
\begin{eqnarray}
&
\widehat{I_1}(p)= 2 \pi P\int_{0}^{\pi}d\theta \sin^{d-2}\theta
\int_{0}^{\infty}dq\cos\theta (\cos\theta-q)q^{2z-3}
\widehat{g_R}\left(p r^{z}\right)\widehat{n}(p q^{z}) \,,
\nonumber
\\
&
\widehat{I_2}(p)= \pi P\int_{0}^{\pi}d\theta \sin^{d-2}\theta
\int_{0}^{\infty}dq(\cos\theta-q)^2 q^{2z-3}r^{-(d+4-2z)}
\widehat{n}\left(p r^{z}\right)\widehat{n}(p q^z) \,.
\nonumber
\end{eqnarray}
%%%%%%%%%%%%%%%%%%%%%%%%%%%%%%%%%%%%%%%%%%%%%%%%%%%%%%%%%%%%%%%%%%
%%%%%%%%%%%%%%%%%%%%%%%%%%%%%%%%%%%%%%%%%%%%%%%%%%%%%%%%%%%%%%%%%%
%%%%%%%%%%%%%%%%%%%%%%%%%%%%%%%%%%%%%%%%%%%%%%%%%%%%%%%%%%%%%%%%%%
%%%%%%%%%%%%%%%%%%%%%%%%%%%%%%%%%%%%%%%%%%%%%%%%%%%%%%%%%%%%%%%%%%
%%%%%%%%%%%%%%%%%%%%%%%%%%%%%%%%%%%%%%%%%%%%%%%%%%%%%%%%%%%%%%%%%%
%%%%%%%%%%%%%%%%%%%%%%%%%%%%%%%%%%%%%%%%%%%%%%%%%%%%%%%%%%%%%%%%%%
\begin{multicols}{2}
\noindent

We now make for $\widehat{n}$ and $\widehat{g}$ the following ansatz:
\begin{eqnarray}
&
\widehat{g}(p)=2C \,\theta(p) \,\exp(-\mid D p\mid^{2/z}) \,,
\label{gansatz}
\\
&
\widehat{n}(p)=A\exp(-\mid B p\mid^{\beta/z}) \,,
\label{nansatz}
\end{eqnarray}
where $A, B, C, D$ are parameters depending on $d$, $z$ and $\lambda$. It 
was actually found that the direct numerical solution of the MC equations in 
one dimension were well-fitted by this ansatz \cite {Hwa}. The same choice 
for $\widehat{g}(p)$ (but not for $\widehat{n}(p)$) was also made in 
\cite{Do}. 
The invariance of Eqs. (\ref{ghat}) and (\ref{nhat}) under the
rescaling: $\widehat{n}(p)\rightarrow \widehat{n}(p/\alpha)/\alpha^2$,
$\widehat{g}(p)\rightarrow \widehat{g}(p/\alpha)$ \cite{Tu}
allows us to choose $D=1$.
This ansatz gives in $x$ space, for large $x$:
\begin{eqnarray}
&
g(x)\simeq const \,x^{-1-2/z}+ i \,\,2C x^{-1} \,,
\nonumber
\\
&
n(x)\simeq const \,x^{-1-\beta/z} \,,
\nonumber
\end{eqnarray}
\noindent
which have the right asymptotic properties. (This seems to be the key 
ingredient in achieving the integer value of $4$ for $d_c$ rather than a 
number like $3.6...$). We are forced to set $C=1/2$ to get the right large $x$ 
behavior of $g_I$. 
Then $\widehat{g_R}$ is simply $\frac{1}{2}e^{-\mid p \mid^{2/z}}$.
We choose to match Eqs. (\ref{ghat}) and (\ref{nhat}) in the limit
$p \rightarrow 0$, which means  matching the most divergent terms
on both sides of these two equations, and is equivalent to matching the
large $x$ behavior.
Since $\mid g(x)\mid^{-2} \simeq x^2$ for large $x$,
in the small $p$ limit, the left hand side in both equations
(\ref{ghat},\ref{nhat}) is dominated by the terms
$d^2\widehat{n}/dp^2$,
$d^2\widehat{g_R}/dp^2$, giving \cite{note}:
\begin{eqnarray}
&
(2-z)/z^2= \lim_{p\rightarrow 0}\mid p\mid^{2-2/z}\widehat{I_1}(p) \,,
\label{l1}
\\
&
AB^{\beta/z}\beta(\beta-z)/z^2 =
\lim_{p\rightarrow 0}\mid p\mid^{2-\beta/z}\widehat{I_2}(p) \,.
\label{l2}
\end{eqnarray}
\noindent
Performing the integrals and taking the limit one gets the following
equations:
\begin{equation}
\frac{P A S_d}{B}=\frac{d (2-z)}{\pi z^2}\frac{1}{B \,I(B,z)} \,,
\label{e1}
\end{equation}
\begin{equation}
\frac{P A S_d}{B^{2}}=\frac{\beta (\beta-z)}{\pi z^2}
\frac{2^{(2z-\beta)/\beta}}{\Gamma(\frac{2z-\beta}{\beta})/\beta} \,,
\label{e2}
\end{equation}
where $S_d=\int_{0}^{\pi}d\theta \sin^{d-2}\theta= 2^{d-2}
\Gamma((d-1)/2))^2/\Gamma(d-1)$
and $I(B,z)\!=\!\int_{0}^{\infty}\!ds (1-2s^2)s^{2z-3}
\exp(-B^{\beta/z}s^{\beta}-s^2)$.
Setting $d=1$, $z=3/2$ in the expression for $B=B(z,d)$ obtained
from Eqs. (\ref{e1}) and (\ref{e2}) we find that $B$ approaches $1$,
and then (since $\beta \rightarrow 2$)
$\widehat{n}(p)\propto \widehat{g_R}(p)$, as required by the
fluctuation-dissipation theorem. In the limit
$d \rightarrow 0$ we get the correct value $z=4/3$.
As $z$ approaches $2$, $B \,I(B,z)\rightarrow \frac{1}{4\sqrt{\pi}}$
and the right hand side of Eq. (\ref{e1}) goes to zero.
It is reasonable to assume that $A$ is a finite number \cite{A},
which forces $B\simeq (2-z)^{-1}$. From comparison with equation
(\ref{e2}) one gets
$2z-\beta=4-d-4(2-z) \simeq (2-z)^2$ which gives $4-d=4(2-z)+O((2-z)^2)$.
Note that this result, not only fixes $d_c$ to be $4$, but also
provide an expansion for $z$ around $d_c^{-}$: $z=2-(4-d)/4$.
In the simplest scenario $B=(const (2-z))^{-1}$,
where the constant is fixed to $2$ by the value of $B$
in $d=1$. Eqs. (\ref{e1}) and (\ref{e2}) are then easily solved
numerically for $1<d<4$,
giving $z\simeq\!1.62$ in $d=2$ and $z\simeq\!1.78$ in $d=3$.
The result in $d=2$ is in good agreement with the values
found in the literature from simulations (which of course do not use the MC
approximation) and also with the direct
numerical solution of the MC equations by Tu, which we suspect may be
accurate for $d\lesssim 2$.

The results $d_c=4$ and $z=2-(4-d)/4+O((4-d)^2)$ are very robust:
assuming a simple scaling form for $\widehat{n}(p)$ as $z$
approaches $2$, they can be derived without any specific assumption
on the form of the scaling functions,
other than the following weak
requirements for the small $p$ behaviour
of the scaling functions:
$\widehat{g_R}(p)\simeq C(1-k p^{2/z})$,
$\widehat{n}(p)\simeq A(1-k'(pB)^{\beta/z})$,
where $k$ and $k'$ are constants of order $1$ as $z\rightarrow 2$.
The scaling form for $\widehat{n}(p)$ is:
\begin{equation}
\widehat{n}(p)=A \widehat{f}(Bp)\,,
\label{scaling}
\end{equation}
where $B=(2(2-z))^{-1}$, $\widehat{f}(0)=1$, and $A f(0)$ is finite
as $z\rightarrow 2$ \cite{Abis} ($f(x)$ denotes the inverse Fourier
transform of $\widehat{f}(p)$).
We fix the scale on which $\widehat{g}$
varies to be of order $1$ \cite{scales}.
When $z$ is close to $2$, $\widehat{g_R}(p)$ is a
slowly varying function compared with $\widehat{n}(p)$, which allows us to
perform the integral $\widehat{I_1}(p)$ to the leading order in $(2-z)$.
Matching Eq. (\ref{ghat}) at $p=0$ gives:
\begin{equation}
\frac{PAS_d}{B}=\frac{d (2-z)}{\pi z^2} \frac{2 k z}{f(0)}.
\label{f1}
\end{equation}
Thus $B\simeq (2-z)^{-1}$, consistent with our assumption.
A similar analysis of Eq. (\ref{nhat}) in the limit
$p\rightarrow 0$ gives
\begin{equation}
\frac{PAS_d}{B^2}=\frac{\beta (\beta-z)}{\pi z^2} (2z-\beta)k \,,
\label{f2}
\end{equation}
which again leads to $B^2\simeq (4z-d-4)^{-1}$, and thus to $d_c=4$,
$z=2-(4-d)/4+O((4-d)^2)$. Eqs. (\ref{f1}) and (\ref{f2}) are of course
consistent with Eqs. (\ref{e1}) and (\ref{e2}) obtained from our 
original ansatz.

Let us now briefly discuss what could have gone wrong in Tu's numerical
analysis. Assume our ansatz (\ref{gansatz}) for $\widehat{g}(p)$
is reliable for all $p$ as $d$ approaches $4$.
Then in $d=4$ $g(x)=(g(0)^{-1}-ix)^{-1}$.
Thus the imaginary part of $g$ has a peak at $x= g(0)^{-1}$.
We broke the scale invariance by choosing $D=1$, which gives $g(0)=1$ and
lead to
$n(0)\simeq (2-z)$. Tu chose instead to break the invariance by
fixing the value of $n(0)$ to be some constant.
Rescaling our solution to make it compatible with Tu's choice, we would
get $g(0)\simeq (2-z)$, which causes the peak in the imaginary part to
run to infinity, making it very difficult to get an accurate evaluation of
the integral $I_1(x)$. It would be very interesting to repeat Tu's analysis
in the light of our findings on the scaling form of $n(x)$ near the upper
critical dimension.

The validity of our results beyond the
mode-coupling approximation is an important consideration. We expect that
the
result $d_c=4$ will hold beyond mode-coupling, as argued in Ref. \cite{Mo}.
However, we know of no argument that would imply that $z\simeq 2-(4-d)/4$
would remain true outside the MC approximation i.e. there might be
corrections at $O(4-d)$ to the value of the dynamical exponent $z$ from the
diagrams not included in the MC approximation.

To summarise, we have  developed a self-consistent approach to the strong
coupling regime of the KPZ equation. We constructed an ansatz for
the scaling functions which is consistent with known asymptotic results,
and gives naturally an integer upper critical dimension $d_c=4$.
It also provides for the first time a satisfactory expansion of $z$ around
$d_c$.
Solving for $z=z(d)$ gives good agreement with previous studies for
all $1\leq d \leq 2$. We showed also
how the same results can be obtained with much weaker requirements
on the form of the scaling functions, if one assumes a scaling form
for $\widehat{n}$.
We found a possible mechanism for the failure of Tu's numerical analysis
as $d$ approached $d_c$.
Our results together with previous numerical MC studies for $d=1$ and $d=2$,
\cite{Hwa,Tu},  and
with the analysis above $d=4$ of Ref. \cite{Mo}, give a semi-quantitative
picture of the behavior of the strong-coupling regime of the KPZ equation
for all dimensions $d$.

The authors acknowledge  the support of EPSRC under grants GR/L38578 and
GR/L97698. F.C. would like to thank Alan Bray, Andrea Cavagna,
Alessandro Flammini, and Uwe T\"auber for discussions.

\end{multicols}

\begin{references}
\bibitem{note} Note that matching the most divergent terms in
the derivatives of Eqs. (\ref{ghat}) and (\ref{nhat})
as $p\rightarrow 0$ gives no new information,
and matching next terms would involve contributions other than the
one coming from $d^2\widehat{n}/d^2p$ and $d^2\widehat{g_R}/d^2p$ in
the left hand side of equation (\ref{ghat},\ref{nhat}).
\bibitem{A}Matching Eq. (\ref{n})
at $x=0$ with our ansatz for $\widehat{n}(p)$
gives $A P =2 /J$ where $J$ is a convergent integral
in the limit $z\rightarrow 2$, $d \rightarrow 4$.
\bibitem{Abis} Again, matching Eq. (\ref{n})
at $x=0$ gives $PAf(0)=4\pi/J'$, where $J'$ is a convergent
integral in the limit $z\rightarrow 2$, $d \rightarrow 4$,
dependent on the specific form of the scaling function $\widehat{f}$.
\bibitem{scales} Due to the rescaling invariance of Eqs.
(\ref{g}) and (\ref{n}), the only relevant quantity is
the relative scale of variation of $\widehat{n}$ and $\widehat{g}$.
\end{references}
\end{document}